\documentclass{amsart}
\usepackage[english]{babel}
\usepackage{amssymb}
\usepackage{graphicx}
\usepackage{epsfig}

\usepackage{amsmath,amsthm,verbatim}
\usepackage{amsrefs}

\newtheorem{theorem}{Theorem}

\theoremstyle{definition}
\newtheorem{defi}[theorem]{Definition}

\theoremstyle{remark}
\newtheorem{remark}[theorem]{Remark}

\newcommand{\dd}{\mathrm{d}}
\newcommand{\barX}{\overline{X}}
\newcommand{\barY}{\overline{Y}}
\newcommand{\f}[2]{\frac{#1}{#2}}

\begin{document}
\title[Semi-Closed Form Cubature]{Semi-Closed Form Cubature and Applications
  to Financial Diffusion Models} 
\author{Christian Bayer, Peter Friz, Ronnie Loeffen}
\address{TU Vienna (first author), TU\ Berlin (second author), WIAS Berlin
  (second and third author)}

\begin{abstract}
  Cubature methods, a powerful alternative to Monte Carlo due to
  Kusuoka~[Adv.~Math.~Econ.~6, 69--83, 2004] and
  Lyons--Victoir~[Proc.~R.~Soc.\\Lond.~Ser.~A 460, 169--198, 2004], involve the
  solution to numerous auxiliary ordinary differential equations. With focus
  on the Ninomiya-Victoir algorithm~[Appl.~Math.~Fin.~15, 107--121, 2008],
  which corresponds to a concrete level $5$ cubature method, we study some
  parametric diffusion models motivated from financial applications, and
  exhibit structural conditions under which all involved ODEs can be solved
  explicitly and efficiently. We then enlarge the class of models for which
  this technique applies, by introducing a (model-dependent) variation of the
  Ninomiya-Victoir method. Our method remains easy to implement; numerical
  examples illustrate the savings in computation time.
\end{abstract}

\keywords{Ninomiya--Victoir method, cubature method, Monte Carlo simulation}
\maketitle

\vspace{0.5cm}

\section{Introduction}

We deal with the common problem in quantitative finance to compute, as fast and
accurately as possible,%
\begin{equation}
\mathbb E\left[ f\left( X_{T}\right) \right] .  \label{EPayoffXT}
\end{equation}%
Here, $f:\mathbb R^N\rightarrow\mathbb R$ denotes a typical payoff function and $\left( X_{t}\right)_{0 \le t \le T} $ is an $N$-dimensional diffusion
process, given in terms of a stochastic differential equation (SDE) in
Stratonovich form
\begin{equation*}
  \begin{pmatrix}
    X_1(t,x) \\
    \vdots \\
    X_N(t,x)
  \end{pmatrix}
  =
  \begin{pmatrix}
    x_1 \\ \vdots \\ x_N
  \end{pmatrix}
  +
  \begin{pmatrix}
    \int_0^t V^1_0(X(s,x))\mathrm{d}s \\ \vdots \\ \int_0^t
    V^N_0(X(s,x))\mathrm{d}s 
  \end{pmatrix}
  + 
  \begin{pmatrix}
    \sum_{j=1}^d \int_0^t V^1_j(X(s,x))\circ \mathrm{d}B^j_s \\ \vdots \\
    \sum_{j=1}^d \int_0^t V^N_j(X(s,x))\circ \mathrm{d}B^j_s 
  \end{pmatrix}
  .
\end{equation*}
where $x=(x_1,\ldots,x_N)\in \mathbb{R}^{N}$ and $B=(B^{1},\ldots ,B^{d})$ is a $d$%
-dimensional standard Brownian motion. Whenever convenient, we shall use the compact notation 
\begin{equation}
X(t,x)=x+\int_{0}^{t}V_{0}(X(s,x))\mathrm{d}s+\sum_{j=1}^{d}%
\int_{0}^{t}V_{j}(X(s,x))\circ \mathrm{d}B_{s}^{j}, \label{initialSDE}
\end{equation}%
or, in It\^{o} form,  
\begin{equation*}
X(t,x)=x+\int_{0}^{t}\widetilde{V}_{0}(X(s,x))\mathrm{d}s+\sum_{j=1}^{d}%
\int_{0}^{t}V_{j}(X(s,x))\mathrm{d}B_{s}^{j},
\end{equation*}%
where $\tilde{V}_{0}^{i}(x)=V_{0}^{i}(x)+\frac{1}{2}\sum_{j=1}^{d}\sum_{k=1}^N V_{j}^{k}\partial _{k}V_{j}^{i}(x)$.%

As is common in the analysis of higher-order, weak
approximation methods for such SDEs (cf. the classics Kloeden and
Platen~\cite{klo/pla92}, Glasserman~\cite{gla04} as well as
Kusuoka~\cite{kus04}, Lyons and Victoir~\cite{lyo/vic04} and Ninomiya and
Victoir~\cite{ninomiyavictoir} for cubature type methods) we shall assume that
the payoff function $f$ and all vector fields $V_{0},V_{1},\dots ,V_{d}$ are
smooth, with bounded derivatives of any order. The standing remark in this
subject, \textit{implicit in all of the aforementioned references}, is that
any scheme obtained from such an analysis can and will be applied to typical
financial diffusion models (such as Heston, SABR and their -- possibly
higher-dimensional -- generalizations) even if they do not satisfy the
technical assumptions initially used in the analysis; numerical experiments
(which are necessary for every numerical scheme in any case!) serve as
a posteriori justification.\footnote{%
  It is possible to analyze mollified/truncated versions of CIR, Heston, SABR,
  \ldots and thus provide further mathematical justification. For instance,
  it was only recently shown in full rigor that the classical Euler-Maruyama
  scheme applied to the Heston model converges; see e.g. Mao and Higham ~\cite{HM05}. Let us also mention the work of Alfonsi \cite{AA10} in this context. Such considerations are not the purpose of the present paper.}

We do not wish to impose any special structure on~\eqref{initialSDE}; in
particular the vector fields are not supposed to commute (cf.~Kloeden and
Platen~\cite{klo/pla92}[page 348] for the advantages in such a case in the
particular case of the Milstein scheme), no affine structure (as in the Heston
model) is assumed, nor do we want to rely on heat-kernel based expansions of
\eqref{EPayoffXT} (such as the SABR formula).  In this generality, one has
essentially two approaches. The \textit{PDE method}, based on the Feynman-Kac
formula, consists in solving the Cauchy problem for the partial differential
equation
\begin{equation*}
  \partial _{t}u\left( t,x\right) + Lu(t,x) = 0,\quad u\left( T,x\right) =
  f(x)
\end{equation*}
where the $2$nd order differential operator $L$ is given in H\"{o}rmander form $L=V_{0}+
\frac{1}{2}\sum_{i=1}^{d}V_{i}^{2}$  where vector-fields are identified with first
order differential operators. As is well known, that PDE approach is
prohibitively slow in higher dimension; there are also stability issues when
$L$ is not elliptic. The other approach is \textit{the probabilistic
  ``simulation'' method} which requires two steps. In \textbf{step 1} one
discretizes $X\left( t,x\right) $ in order to obtain an approximation $%
\barX^{K}\left( t,x\right) $; typically, $K$ corresponds to the number of
partitions of $\left[ 0,T\right] $; examples include the Euler-Maruyama (EM)
scheme
\begin{eqnarray*}
  \barX^{\left( EM\right) ,K}\left( 0,x\right) &=&x \in\mathbb R^N \\
  \barX^{\left( EM\right) ,K}\left( \frac{k+1}{K},x\right) &=&\barX^{\left(
      EM\right) ,K}\left( \frac{k}{K},x\right)
  +\widetilde{V}_{0}(\barX^{\left( EM\right) ,K}\left( \frac{k}{K},x\right)
  )\mathrm{\times }\frac{T}{K} \\ 
  &&+\sqrt{\frac{T}{K}}\sum_{j=1}^{d}V_{j}\left( \barX^{\left( EM\right) ,K}\left( 
      \frac{k}{K},x\right) \right) Z_{k+1}^{j},
\end{eqnarray*}%
where $\left( Z_{k}^{j}\right) $ is a family of independent $\mathcal{N}(
0,1) $\footnote{Throughout the paper $\mathcal{N}(\mu,\sigma^2)$ denotes the normal distribution with mean $\mu$ and variance $\sigma^2$.} random variables, as well as higher order (Milstein, Kusuoka,
Ninomiya--Victoir, \ldots) schemes which we do not wish to detail at this
moment. The \textit{discretization error} is given by%
\begin{equation*}
  \left\vert \mathbb E\left[ f\left( X\left( T,x\right) \right) \right] - \mathbb E\left[
      f\left( \barX^{K}\left( T,x\right) \right) \right] \right\vert =\left\{ 
    \begin{array}{c}
      \mathcal{O}\left( T/K\right) \text{ for Euler-Maruyama} \\ 
      \mathcal{O}\left( (T/K)^{2}\right) \text{ for Ninomiya-Victoir\ } \\ 
      \cdots%
    \end{array}%
  \right.
\end{equation*}%
In \textbf{step 2} one has to integrate $ f\left( \barX^{K}\left(
T,x\right) \right)$ over some domain of dimension $D=D\left(
K\right) $ such as\footnote{%
The dimension $D(K) $ will depend on the method (for instance $%
D(K) =K\times d$ for the Euler-Maruyama scheme, $D(K) =K\times ( d+1) $ for
the Ninomiya--Victoir scheme).}% 
\begin{equation*}
  \mathbb E\left[ f\left( \barX^{K}\left( T,x\right) \right) \right]
  =\int_{[0,1)^{D\left( K\right) }}F\left( y_{1},\dots ,y_{D\left( K\right)
    }\right) dy_{1}\dots dy_{D\left( K\right) }.
\end{equation*}%
Here, $F$ denotes the dependence of $f\left( \barX^{K}\left( T,x\right)
\right)$ on uniform random variables, i.e., $F\left( U_{1},\dots ,U_{D\left(
      K\right) }\right) = f\left( \barX^{K}\left( T,x\right) \right)$ for a
collection $\left( U_{1},\dots ,U_{D\left( K\right) }\right)$ of independent
random variables uniformly distributed on the unit interval. The
right-hand-side is approximated by Monte Carlo (MC) or Quasi Monte Carlo
(QMC), essentially obtained by averaging $M$ samples of $F\left( y_{1},\dots
  ,y_{D\left( n\right) }\right) $. These samples are random if created by
Monte Carlo (MC) and deterministic if obtained by Quasi Monte Carlo (QMC).  In
either case, we have an \textit{integration error }of the form%
\begin{eqnarray*}
  &&\left\vert \mathrm{MC}\left( f\left( \barX^{K}\left( T,x\right) \right)
      ,M\right) \left( \omega \right) - \mathbb E\left[ f\left( \barX^{K}\left(
          T,x\right) \right) \right] \right\vert, \\
  &&\left\vert \mathrm{QMC}\left( f\left( \barX^{K}\left( T,x\right) \right)
      ,M\right) - \mathbb E\left[ f\left( \barX^{K}\left( T,x\right)
      \right) \right] \right\vert.
\end{eqnarray*}%
The central limit theorem roughly implies that $\mathrm{MC}$-integration
error is $O\left( 1/\sqrt{M}\right) $. More precisely, we have in the sense of
an asymptotic equality in law,
\begin{equation*}
\mathrm{MC}\left( f\left( \barX^{K}\left( t,x\right) \right) ,M\right) \approx
\mathcal{N}\left( \mathbb{E}\left[ f\left( \barX^{K}\left( t,x\right) \right)
  \right] , \mathbb{V}\left[ f\left( \barX^{K}\left( t,x\right) \right)
  \right] /M\right)
\end{equation*}
so that, using $\mathbb{V}\left[ f\left( \barX^{K}\left( t,x\right) \right) %
\right] \approx \mathbb{V}\left[ f\left( X\left( t,x\right) \right) \right] $
we see that the number of sample points $M$ needed to attain a given accuracy
(i.e. a certain $\varepsilon $ bound for the $\mathrm{MC}$-integration error)
is roughly independent of $K$ and the discretization algorithm. The situation is
somewhat different for the $\mathrm{QMC}$-integration error. It is known that
there exists sequences ("sample points") such that there exists $%
C=C\left( f,D\left( K\right) \right) $ such that for all $M$ one has%
\begin{equation*}
\left\vert \mathrm{QMC}\left( f\left( \barX^{K}\left( T,x\right) \right)
,M\right) \left( \omega \right) - \mathbb E\left[ f\left( \barX^{K}\left( T,x\right)
\right) \right] \right\vert \leq C\frac{\left( \log M\right) ^{D\left(
K\right) }}{M}.
\end{equation*}%
In contrast to the MC case, the number of sample points $M$ needed by QMC to
attain a given accuracy depends heavily on the dimension of integration $%
D\left( K\right) $ and, possibly, on the smoothness of $f\left(\barX^{K}
\right)$ as a function in the points $y_1, \ldots, y_{D(K)}$. Moreover, the
above error estimate is known to grossly overestimate the true error in many
cases. 

\subsection{Cubature on Wiener Space}

Let us briefly put the (Kusuoka--Lyons--Victoir) cubature method in this
context. For simplicity of notation only, we consider the case $V_{0}=0$
here. A \textit{cubature formula on Wiener space }is a random variable $W$
taking values in the space $C^{\text{1-var}}([0,1],\mathbb{R}^{d})$ of
continuous paths of bounded variation with values in $\mathbb{R}^{d}$ such
that we have
\begin{equation}
\mathbb{E}\left[ \int_{0\leq t_{1}\leq \cdots \leq t_{j}\leq 1}\circ
\mathrm dB_{t_{1}}^{i_{1}}\cdots \circ \mathrm dB_{t_{j}}^{i_{j}}\right] =\mathbb{E}\left[
\int_{0\leq t_{1}\leq \cdots \leq t_{j}\leq 1}\mathrm dW_{t_{1}}^{i_{1}}\cdots
\mathrm dW_{t_{j}}^{i_{j}}\right] .  \label{eq:cubature-formula}
\end{equation}%
for all multi-indices $I=(i_{1},\ldots ,i_{j})\in \{1,\ldots ,d\}^{j}$ with  all $1\leq j\leq m$, where $m$ is a fixed positive integer, the
\textit{order }of the cubature formula. Moreover, we note that since the paths
of the process $W$ are of bounded variation, the integrals on the right hand
side of~(\ref{eq:cubature-formula}) are then understood as classical
Riemann-Stieltjes integrals. In applications, the reference interval $\left[
  0,1\right] $ in (\ref{eq:cubature-formula}) is typically replaced by some
(small)\ interval such as $\left[ 0,T/K\right] $. (Due to Brownian scaling,
however, the problems are equivalent; in particular, a cubature formula on $
\left[ 0,t\right] $ is obtained by a scaled version of the cubature formula
on $\left[ 0,1\right] $.) In the classical paper of Lyons and
Victoir~\cite{lyo/vic04} the authors actually insisted that the cubature
formula is \textit{discrete} meaning that for some positive integer $k$, the law of $W$ can be written as
\begin{equation*}
\sum_{i=1}^{k}\lambda _{i}\delta _{W_{i}},
\end{equation*}
where $\delta _{W_{i}}$ is the Dirac measure on Wiener space which assign
unit mass to the path $W_{i}\left( .\right) ,$ zero to every other path.
(Existence and explicit knowledge of cubature formulas is a non-trivial
problem!) The idea is now to approximate the stochastic differential equation
\eqref{initialSDE} for $X=X\left( t,x\right) $ by a family of (random)
ordinary time-inhomogeneous differential equations,
\begin{equation*}
\barX(t,x;W)=x+\sum_{j=1}^{d}\int_{0}^{t}V_{j}(\barX(s,x;W))
\frac{\dd W^{j}\left( s\right) }{\dd s} \dd s,
\end{equation*}
where $W$ now denotes a cubature formula on the interval $[0,t]$.
A stochastic Taylor-expansions (e.g. chapter 18 in \cite{friz-victoir-book} 
for a discussion in the spirit of cubature) shows that $\mathbb{E}\left[ f\left(
    \barX(t,x;W)\right) \right] -\mathbb{E}\left[ f\left( X\left( t,x\right)
  \right) \right] =O\left( t^{\frac{m+1}{2}}\right) $ as $t\rightarrow 0.$
Observe that in the case of a discrete cubature formula $\mathbb{E}\left[
  f\left( \barX(t,x;W)\right) \right] $ is computed exactly (no integration
error!) by solving $k$ ordinary differential equations. A (big) interval
$\left[ 0,T \right] $ can be handled by dividing it into $K$ intervals of
length $T/K$ and iterating this procedure but now exact computation of
$\mathbb{E}\left[ f\left( X(t,x;W)\right) \right] $ requires to solve
\begin{equation*}
k+k^{2}+\dots +k^{K}=O\left( k^{K}\right)
\end{equation*}%
ordinary differential equations. When $k^{K}$ becomes too big one can either
perform a Monte Carlo simulation (``on the cubature tree'') or resort to
recombination techniques (see Litterer and Lyons~\cite{lyo/lit10} for the
present state of art). Let us note, however, that in many practical
applications $K$ remains small, which helps to explain the numerical benefits
of cubature even without recombination.

\subsection{The Ninomiya--Victoir (NV) Scheme}\label{sectionNVscheme}

The Ninomiya--Victoir ``splitting'' scheme, introduced
in~\cite{ninomiyavictoir}, is given by
\begin{align*}
\barX^{\left( NV\right) ,K}&\left( 0,x\right) = x\in \mathbb R^N, \\
\barX^{\left( NV\right) ,K}&\left( \frac{(k+1)T}{K},x\right) =\\
&=\left\{ 
\begin{array}{c}
\mathrm e^{\frac{T}{2K}V_0}\mathrm e^{Z^1_k \sqrt{\f{T}{K}} V_1}\cdots \mathrm e^{Z^d_k
  \sqrt{\f{T}{K}} V_d} \mathrm e^{\frac{T}{2K} V_0} \barX^{\left( NV\right) ,K}\left( 
\frac{kT}{K},x\right) \text{ \ \ if }\Lambda _{k}=-1, \\ 
\mathrm e^{\frac{T}{2K}V_0} \mathrm e^{Z^d_k \sqrt{\f{T}{K}} V_d}\cdots \mathrm e^{Z^1_k
  \sqrt{\f{T}{K}} V_1} \mathrm e^{\frac{T}{2K} V_0} \barX^{\left( NV\right) ,K}\left( 
\frac{kT}{K},x\right) \text{ \ \ if }\Lambda _{k}=+1.%
\end{array}
\right.
\end{align*}
Here $\mathrm e^{V}x\in\mathbb R^N$ denotes the ODE\ solution at unit time to $\dot{y}=V\left(
  y\right) ,y\left( 0\right) =x$ and the probability space carries independent
random-variables $\left( \Lambda _{k}\right) $, with values $\pm 1$ at
probability $1/2$, and $\mathcal{N}\left( 0,1\right) $ random variables
$(Z_{k}^{j})$.  One step in the NV scheme corresponds actually to a
(non-discrete) cubature formula of order $m=5$. To see this, assume $V_{0}=0$
for (consistent) simplicity and let $\left( \mathfrak{b}_{i}\right) $ denote
the canonical basis of $\mathbb{R}^{d}$. An $\mathbb{R}^{d}$-valued random
path $W\left( \omega \right) $, continuous and of bounded variation, is then
created via $ \Lambda \left( \omega \right) \in \left\{ +1,1\right\} $ and $d$
independent $\mathcal{N}\left( 0,1\right) $ realizations $Z^{1}\left( \omega
\right) ,\dots ,Z^{d}\left( \omega \right) $. If $\Lambda \left( \omega
\right) =-1$ we take $W\left( \omega \right) :\left[ 0,T/K\right] \rightarrow
\mathbb{R}^{d}$ , started at $0$ say, to move at constant speed, first an
amount $Z_{k}^{d} \sqrt{T/K}$ in $\mathfrak{b}_{d}$-direction, \ldots until
the final move $ Z_{k}^{1} \sqrt{T/K}$ in $\mathfrak{b}_{1}$-direction; if
$\Lambda \left( \omega \right) =+1$ the construction is similar but in
reversed order. When $ V_{0}\neq 0$, one follows the flow of the drift
vector-field for time $ T/\left( 2K\right) $ in the first and last step of the
scheme; at all intermediate steps $V_{0}$ is followed for a time $T/K$; this
is inspired by classical splitting methods in operator theory. Let us also
note that the coin-flipping corresponds to Talay's trick of, in a weak
approximation context, replacing the (difficult to sample) L\'{e}vy's area by
a discrete moment-matched random variable, see Kloeden and Platen~\cite{klo/pla92}[page
466 f.]. 

The NV scheme has attracted wide attention since its introduction in ~\cite{ninomiyavictoir}; it is nowadays found in various sophisticated numerical packages such as Inria's software PREMIA for financial option computations. \footnote{As of Sep 2010, the weblink ralyx.inria.fr/2006/Raweb/mathfi/uid21.html contains some relevant information.} A variation of the scheme designed to deal with degeneracies arising some affine situations is discussed in ~\cite{AA10}. Let us also mention the "NV inspired" schemes 
developed in ~\cite{fuj} and ~\cite{otv09}. 

\subsection{Semi-closed form cubature}

It is clear from the preceding discussion that cubature methods, and the NV
scheme in particular, heavily rely on the ability to solve, fast and
accurately, ordinary differential equations. The general cubature methods
involves \textit{time-inhomogeneous }ODEs; in general, there is no
alternative to solve them numerically, typically with Runge-Kutta methods.
(A detailed discussion on how Runge-Kutta methods are applied in this
context is found in Ninomiya and Ninomiya~\cite{nin/nin09}.)

On the other hand, the Ninomiya-Victoir splitting scheme only involves the
composition of solution flows to \textit{time-homogeneous} ODEs. In particular,
there will be "lucky" cases of models where all\ (or at least most) ODE\ flows
can be solved exactly.\footnote{
  By this we mean a closed-form solution to the ODE\ $\dot{y}=V\left( y\right)
  ,y\left( 0\right) =x$ \textit{which allows for fast numerical evaluation}.
  In particular, we are not interested in "closed-form" solution in terms of
  complicated and slow-to-evaluate special functions.} In such a case one has
effectively found a level-$5$ cubature method which can be implemented without
relying on numerical ODE\ solvers. In particular, one expects the cubature
methods to perform especially well in such cases. As was observed
in~\cite{ninomiyavictoir}, see also Section \ref{example_heston}, the Heston model is such a lucky
case. We thus propose the following definition.

\begin{defi}
A diffusion model of type (\ref{initialSDE}) where a cubature method can be
implemented without any numerical ODE\ solutions is said to be accessible to 
\textit{semi-closed form cubature (SCFC)}.
\end{defi}

For instance, any model of type (\ref{initialSDE}) where all ODE flows
$\mathrm e^{tV_{0}},\dots ,\mathrm e^{tV_{d}}$ can be solved in closed form falls in this
class. However, one soon encounters model (e.g. the popular SABR model, see
Section~\ref{sec:example:-sabr-model}) in which some of the vector-fields do
not allow for flows in closed form. The contribution of this paper,
\textit{beyond suggesting the systematic use of financial models that are
  accessible to semi-closed form cubature}, is that the class of such models
can be significantly enlarged by working with an almost trivial modification of
the NV scheme. \footnote{While SCFC corresponds to the ``luckiest'' case of avoiding numerical ODE solvers altogether, any significant reduction of numerical ODEs to be solved will be desirable. Our modification of the NV scheme can obviously be used to this purpose as well.}
 Before explaining our modification we point out that the SABR
model then becomes accessible to semi-closed form cubature. Our modification
is based on the trivial equivalence of (\ref{initialSDE}) with
\begin{equation*}
 \begin{split}
dX(t,x) =& \left( V_{0}(X(t,x))- \sum_{j=1}^d \gamma _{j}V_{j}(X(t,x))\right)
\mathrm{d} t+\sum_{j=1}^{d}V_{j}(X(t,x))\circ \mathrm{d}\left( B_{t}^{j}+\gamma
_{j}t\right)    \\
\equiv & V^{(\gamma)}_{0}\left( X(t,x)\right) +\sum_{j=1}^{d}V_{j}(X(t,x))\circ 
\mathrm{d}\left( B_{t}^{j}+\gamma _{j}t\right) 
%\label{initialSDErewrittenWithDrift}
\end{split}
\end{equation*}%
whatever the choice of drift parameters $\gamma _{1},\dots ,\gamma _{d}$.
Assume that all diffusion vector-fields ($V_{1},\dots ,V_{d}$) allow for flows
in closed form, whereas $\mathrm e^{tV_{0}}$ is not available in closed form.  The
point is that, in a variety of concrete examples, one can pick drift
parameters $\gamma_1, \ldots, \gamma_d$ in a way that $\mathrm e^{tV^{(\gamma)}_{0}}$
can be solved in closed form after all.

Therefore, we propose the following variant of the Ninomiya-Victoir method
(which shall be referred to as the ``NV scheme with drift
(trick)''):
\begin{align} 
  \barX^{\left( NVd\right) ,K}&\left( 0,x\right) = x\in\mathbb R^N, \nonumber \\
  \label{eq:NV-drift}
  \barX^{\left( NVd\right) ,K}&\left( \frac{(k+1)T}{K},x\right) =\\
  &=\left\{ 
    \begin{array}{c}
      \mathrm e^{\frac{T}{2K}V^{(\gamma)}_0} \mathrm e^{Z^1_k V_1} \cdots \mathrm e^{Z^d_k V_d}
      \mathrm e^{\frac{T}{2K} V_0^{(\gamma)}} \barX^{\left( NVd\right) ,K} \left( 
        \frac{kT}{K},x\right), \text{ \ \ if }\Lambda _{k}=-1, \\ 
      \mathrm e^{\frac{T}{2K}V_0^{(\gamma)}} \mathrm e^{Z^d_k V_d}\cdots \mathrm e^{Z^1_k V_1}
      \mathrm e^{\frac{T}{2K} V_0^{(\gamma)}} \barX^{\left( NVd\right) ,K}\left(  
        \frac{kT}{K},x\right), \text{ \ \ if }\Lambda _{k}=+1,
    \end{array} \label{eq:NV-driftpart2}
  \right.
\end{align}
where $Z^i_k \sim \mathcal{N}\left(\f{T}{K} \gamma_i,\; \f{T}{K} \right)$
independent of each other.

The bulk of this paper is devoted to implement these ideas for a handful of
(stochastic volatility) models encountered in the financial industry. Since
high-dimensional problems are the raison d'\^{e}tre for probabilistic
simulation methods, a detailed discussion of a higher-dimensional
(SABR-type) model is included. At last, we discuss numerical results
obtained with our ``drift-modified'' NV scheme: relative to the classical
NV scheme we observe significant and consistent savings in computational time.

Note that we want to concentrate on the method itself, without further
improvements like variance reduction, optimization of code and Romberg
extrapolation.

\vspace{0.5cm}

{\bf Acknowledgment:} Partial support of MATHEON and the European Research Council under the European Union's Seventh Framework Programme (FP7/2007-2013) / ERC grant agreement nr. 258237 is gratefully acknowledged. 

\newpage

\section{Application of classical NV scheme to Heston and SABR}

\subsection{Heston model}\label{example_heston}

The stochastic volatility model of Heston is given by the SDE:
\begin{equation*}
\begin{split}
\mathrm{d}X_1(t,x)= & \mu X_1(t,x)\mathrm{d}t +
\sqrt{X_2(t,x)}X_1(t,x)\mathrm{d}B^1_t  \\ 
\mathrm{d}X_2(t,x)= & \kappa(\theta - X_2(t,x))\mathrm{d}t +
\xi\sqrt{X_2(t,x)} \mathrm{d} \left( \rho B_t^1 + \sqrt{1-\rho^2}B_t^2
\right), 
\end{split}
\end{equation*}
where $\mu$ is the rate of return of the asset, $\theta$ is the long vol,
$\kappa$ is the mean-reversion rate, $\xi$ is the vol(atility) of vol(atility)
and $\rho$ is the correlation parameter between the (standard) Brownian
motions $B_1$ and $\left( \rho B_t^1 + \sqrt{1-\rho^2}B_t^2 \right)$.

The   vector fields are given by
\begin{equation*}
\widetilde{V}_0(x)=
\begin{pmatrix}
\mu x_1 \\ \kappa(\theta - x_2)
\end{pmatrix}
,\quad
V_1(x)=
\begin{pmatrix}
\sqrt{x_2} x_1 \\ \xi\rho\sqrt{x_2}
\end{pmatrix}
,\quad
V_2(x)=
\begin{pmatrix}
0 \\ \xi\sqrt{1-\rho^2}\sqrt{x_2}
\end{pmatrix}
\end{equation*}
and so we get
\begin{equation*}
\begin{split}
\begin{pmatrix}
V_0^1(x) \\ V_0^2(x)
\end{pmatrix}
= &
\begin{pmatrix}
\widetilde{V}_0^1(x) \\ \widetilde{V}_0^2(x)  
\end{pmatrix}
-
\begin{pmatrix}
\frac12 \sum_{j=1}^2 V_jV_j^1(x) \\ \frac12 \sum_{j=1}^2 V_jV_j^2(x)
\end{pmatrix}
% =
% \begin{pmatrix}
% \mu x_1 - \frac12 [x_2x_1+\frac12\xi\rho x_1] \\ \kappa(\theta - x_2) -\frac12[\frac12 \xi^2\rho^2 + \frac12 \xi^2(1-\rho^2)]
% \end{pmatrix} \\
= 
\begin{pmatrix}
[\mu-\frac14\xi\rho] x_1 - \frac12 x_2x_1  \\ \kappa(\theta - x_2) -\frac14 \xi^2
\end{pmatrix}. 
\end{split}
\end{equation*}
The corresponding solutions to the ODEs are (cf. Lord et al. \cite{lordetal}*{p.8-9} and their reference to \cite{ninomiyavictoir}; see also the Appendix)
\begin{equation*}
\mathrm{e}^{sV_0}x=
\begin{pmatrix}
 x_1\exp \left( [\mu-\frac14\xi\rho-\frac12J]s +
   \frac12\frac{x_2-J}{\kappa}[\mathrm{e}^{-\kappa s} -1]   \right) \\ 
(x_2-J) \mathrm{e}^{-\kappa s} + J 
\end{pmatrix},
\end{equation*}
\begin{equation*}
\mathrm{e}^{sV_1}x=
\begin{pmatrix}
 x_1 \exp \left( \frac{\left( \frac12 \xi\rho s +\sqrt{x_2}  \right)_+^2
     -x_2}{\xi\rho} \right) \\ 
 \left( \frac12 \xi\rho s +\sqrt{x_2}  \right)_+^2
\end{pmatrix},
\end{equation*}
\begin{equation*}
\mathrm{e}^{sV_2}x=
\begin{pmatrix}
x_1 \\
 \left( \frac12\xi\sqrt{1-\rho^2}s+\sqrt{x_2} \right)_+^2  \\
\end{pmatrix},
\end{equation*}
with $J=\frac{\kappa\theta-\frac14\xi^2}{\kappa}$. We assume (as in \cite{lordetal}) that $J\geq0$; see \cite{AA10} for how to proceed otherwise.

% However, the second elements of $\mathrm{e}^{sV_1}x$ and $\mathrm{e}^{sV_2}x$
% are not the only solutions in the case $x_2=0$. Indeed for $V_1$ instead of
% $f(s)=  (\frac12 \xi\rho s ) _+^2$ we can take $f(s)=0$ or $f(s)= ( \frac12
% \xi\rho s )_+^2\mathbf{1}_{\{s\geq S\}}$, $S>0$. Note that on the negative
% half line, the solution is unique. The choice we have made above, is the same
% one as Lord et al. \cite{lordetal}*{p.8-9} have made. Note the mistake
% regarding the solutions to these ODEs in Ninomiya and Victoir
% \cite{ninomiyavictoir}*{Equation 12}. Further, the NV-scheme will bug if
% $J<0$.

The Heston model can be rewritten in log-coordinates. Define $Y_1(t)=\log X_1(t,x)$ and $Y_2(t)=X_2(t)$. In this new coordinate chart, the vector fields are 
\begin{equation*}
\begin{split}
\begin{pmatrix}
V_0^1(y) \\ V_0^2(y)
\end{pmatrix}
= &
\begin{pmatrix}
[\mu-\frac14\xi\rho] - \frac12 y_2  \\ \kappa(\theta - y_2) -\frac14 \xi^2
\end{pmatrix} 
,\quad
V_1(y)=
\begin{pmatrix}
\sqrt{y_2}  \\ \xi\rho\sqrt{y_2}
\end{pmatrix}
\end{split}
\end{equation*}
and the corresponding solutions to the ODEs are
\begin{equation*}
\mathrm{e}^{sV_0}y=
\begin{pmatrix}
 y_1 + [\mu-\frac14\xi\rho-\frac12J]s +
 \frac12\frac{y_2-J}{\kappa}[\mathrm{e}^{-\kappa s} -1]    \\ 
(y_2-J) \mathrm{e}^{-\kappa s} + J 
\end{pmatrix},
\end{equation*}
\begin{equation}\label{vectorfieldslogheston}
\mathrm{e}^{sV_1}y=
\begin{pmatrix}
 y_1 +  \frac{\left( \frac12 \xi\rho s +\sqrt{y_2}  \right)_+^2 -y_2}{\xi\rho}
 \\ 
 \left( \frac12 \xi\rho s +\sqrt{y_2}  \right)_+^2
\end{pmatrix},
\end{equation}
\begin{equation*}
\mathrm{e}^{sV_2}y=
\begin{pmatrix}
y_1 \\
 \left( \frac12\xi\sqrt{1-\rho^2}s+\sqrt{y_2} \right)_+^2  \\
\end{pmatrix},
\end{equation*}
with $J=\frac{\kappa\theta-\frac14\xi^2}{\kappa}$, as before and $y=(y_1,y_2)=(\log x_1,x_2)$. As is well-known, it is far preferable to use Heston in log-coordinates when simulating with the EM scheme. Although this is less critical in the cubature context, we still recommend \eqref{vectorfieldslogheston} to avoid the numerical evaluation of $\exp(\cdot)$.

\subsection{SABR model}
\label{sec:example:-sabr-model}

The SABR model is given by
\begin{equation*}
\begin{split}
\mathrm{d}X_1(t,x)= &  a  X_2(t,x) (X_1(t,x))^\beta\mathrm{d}B^1_t  \\
\mathrm{d}X_2(t,x)= &  b X_2(t,x) \mathrm{d} \left( \rho B_t^1 +
  \sqrt{1-\rho^2}B_t^2  \right), 
\end{split}
\end{equation*}
where $\frac12\leq\beta\leq1$, $a,b>0$ and $-1<\rho< 1$. \footnote{Although in the literature the SABR model is also considered for $0<\beta<\frac12$ we restrict ourselves to the case $\frac12\leq\beta\leq 1$ in order to avoid difficulties regarding well-posedness of $X$, cf. \cite{lionsmusiela2}.}
The corresponding vector fields are
\begin{equation*}
\widetilde{V}_0(x)=
\begin{pmatrix}
0 \\ 0
\end{pmatrix}
,\quad
V_1(x)=
\begin{pmatrix}
a x_2 x_1^\beta \\ b\rho x_2
\end{pmatrix}
,\quad
V_2(x)=
\begin{pmatrix}
0 \\ b\sqrt{1-\rho^2} x_2
\end{pmatrix}
\end{equation*}
and so we get
\begin{equation*}
\begin{split}
\begin{pmatrix}
V_0^1(x) \\ V_0^2(x)
\end{pmatrix}
= &
-
\begin{pmatrix}
\frac12 \sum_{j=1}^2 V_jV_j^1(x) \\ \frac12 \sum_{j=1}^2 V_jV_j^2(x)
\end{pmatrix}
=
% \begin{pmatrix}
% - \frac12 [\beta x_2^2 x_1^{2\beta-1}+\alpha\rho x_2x_1^\beta ] \\
% -\frac12[\alpha^2\rho^2 x_2 + \alpha^2(1-\rho^2)x_2] 
% \end{pmatrix} \\
% = & 
\begin{pmatrix}
- \frac12 [a^2\beta x_2^2 x_1^{2\beta-1}+ab\rho x_2x_1^\beta ]  \\  -\frac12 b^2 x_2 
\end{pmatrix}. 
\end{split}
\end{equation*}
The solutions to the ODEs corresponding to the vector fields $V_1$ and $V_2$ are
\begin{equation*}
\mathrm{e}^{sV_1}x=
\begin{pmatrix}
 g_1(s) \\
 x_2\exp \left( b  \rho  s \right) 
\end{pmatrix},
\end{equation*}
\begin{equation*}
\mathrm{e}^{sV_2}x=
\begin{pmatrix}
x_1 \\
 x_2\exp \left( b \sqrt{1-\rho^2} s \right)  
\end{pmatrix},
\end{equation*}
where
\begin{equation*}
\begin{split}
 g_1(s) = & \left[ (1-\beta)\frac{a x_2}{b\rho} \left(
     \mathrm{e}^{ b\rho s}-1 \right) +x_1^{1-\beta}
 \right]^{1/(1-\beta)}_+, \quad 0<\beta<1, \\ 
 g_1(s) = & x_1\exp \left( \frac{a x_2}{ b\rho} \left(
     \mathrm{e}^{ b\rho s}-1 \right)  \right)  , \quad \beta=1. 
\end{split}
\end{equation*}
For details on the uniqueness of $g_1$ we refer to the Appendix. Concerning the solution to the ODE corresponding to $V_0$, let $H(s)$ be the first component of $\mathrm{e}^{sV_0}x$, i.e.
\begin{equation*}
\mathrm{e}^{sV_0}x=
\begin{pmatrix}
 H(s) \\
 x_2\exp \left( -\frac12 b^2 s \right) 
\end{pmatrix}.
\end{equation*}
% We need to solve the ODE
% \begin{equation*}
% \begin{pmatrix}
% \dot{z}_1(t) \\ \dot{z}_2(t)
% \end{pmatrix}
% =z'(t)= s V_1(z(t))=
% \begin{pmatrix}
% s z_2 z_1^\beta \\ s\alpha\rho z_2
% \end{pmatrix}
% , \quad (z_1(0),z_2(0))=(x_1,x_2).
% \end{equation*}
% It follows that
% \begin{equation*}
% z_2(t)= x_2\mathrm{e}^{s\alpha\rho t}
% \end{equation*}
%  and assuming $x_1>0$,
% \begin{equation*}
% z_1(t)= \left( (1-\beta)s\int_0^t z_2(u)\mathrm{d}u +x_1^{1-\beta}
% \right)^{1/(1-\beta)} 
% = \left( \frac{1-\beta}{\alpha\rho}x_2 \left( \mathrm{e}^{s\alpha\rho t} -1\\
%  \right)  +x_1^{1-\beta} \right)^{1/(1-\beta)}.
% \end{equation*}
% Note that if $s<0$, then $z_1(t)$ might only be well-defined on part of the
% positive half-line. In that case, let $\tau=\inf\{t>0:z_1(t)=0\}$ and set
% $z_1(t)=0$ for $t>\tau$. 
% We also need to solve the IVP 
% \begin{equation*}
% \begin{pmatrix}
% \dot{z}_1(t) \\ \dot{z}_2(t)
% \end{pmatrix}
% =z'(t)= s V_0(z(t))=
% \begin{pmatrix}
% - \frac12 s[\beta z_2^2 z_1^{2\beta-1}+\alpha\rho z_2z_1^\beta ] \\  -\frac12
% s\alpha^2 z_2 
% \end{pmatrix}
% , \quad (z_1(0),z_2(0))=(x_1,x_2).
% \end{equation*}
% We get
% \begin{equation*}
% z_2(t)= x_2\mathrm{e}^{-\frac12 s\alpha^2 t}
% \end{equation*}
It is impossible to find $H$ in closed-form (unless $\rho=0$  or $\beta=1$). This means that applying
the standard NV-scheme must involve the numerical solution of auxiliary
ODEs. We shall see later that with the NV scheme with drift all involved ODEs
can be solved in closed form.
%Note that when $0<\beta<\frac12$  extra complications arise regarding the existence and uniqueness of $H(s)$, see Appendix????????????, and therefore this case will be excluded from the analysis. 

\section{Models accessible to SCFC and NV with drift}

\subsection{Motivation}

In the classical NV scheme, only centered Gaussian (Brownian) increments are
used to flow along the diffusion vector fields.  Our main observation is that
one can also use non-centered Gaussian increments; this affects the drift term
and, chosen in a smart way, can sometimes render all auxiliary ODE to be
solvable in closed form. To motivate the class of models for which this works,
we illustrate how to systematically construct models accessible to SCFC from  a fairly general two-factor stochastic volatility model given in
It\^o form by 
\begin{equation*}
\begin{split}
\mathrm{d}X_1(t)= &    A(X_1(t))B(X_2(t)) \mathrm{d}B^1_t  \\
\mathrm{d}X_2(t)= &  C(X_2(t))\mathrm{d}t +  D(X_2(t))\mathrm{d}B_t^1 +
E(X_2(t)) \mathrm{d}B_t^2, 
\end{split}
\end{equation*}
where $(X_1(0),X_2(0))=(x_1,x_2)$ is kept fixed. 
In Stratonovich form this becomes (omitting the dependence on $t$ in the drift
and diffusion coefficients), 
\begin{equation*}
\begin{split}
\mathrm{d}X_1(t)= &    -\frac12 A(X_1)\left[ A'(X_1)B^2(X_2) +D(X_2)B'(X_2)
\right] \mathrm{d}t  
+ A(X_1)B(X_2) \circ \mathrm{d}B^1_t  \\
\mathrm{d}X_2(t)= &  \left[ C(X_2)-\frac12 \left[ D(X_2)D'(X_2) +E(X_2)E'(X_2)
  \right]  \right] \mathrm{d}t \\
& +  D(X_2) \circ \mathrm{d}B_t^1 + E(X_2) \circ \mathrm{d}B_t^2, 
\end{split}
\end{equation*}
In the subsequent analysis we shall exhibit a number of possible choices which lead to models accessible to SCFC.
First we would like to choose the coefficients $A,\ldots,E$ such that  we can
rewrite the first SDE as 
\begin{equation*}
\mathrm{d}X_1(t)=    H_1(X_1)H_2(X_2) \mathrm{d}t + A(X_1)B(X_2) \circ \mathrm{d}
\left( B^1_t + \gamma_1 t \right)  
\end{equation*}
% for a constant $\gamma$ and $H$ consisting of one term only. We see that in
% order to do this, we need 
for a constant $\gamma_1$ and functions $H_1,H_2$. Three possible ways
to achieve this goal are
\begin{equation*}
\text{(i) $A'(X_1)\propto 1$, \quad (ii) $D(X_2)=0$ \quad or \quad
  (iii) $D(X_2)B'(X_2)\propto B(X_2)$}. 
\end{equation*} 
Note that the Heston model is a particular example satisfying (i). However, in case (i) we would have $\gamma_1=0$ and since we want to illustrate the additional benefit of the NV scheme with drift over the classical NV scheme, we will not consider case (i) in any more detail. Moreover, since we would like the volatility factor to
depend on the Brownian motion driving the stock, we will also skip case
(ii) and concentrate on case (iii) which implies
$\frac{B(X_2)}{B'(X_2)} \propto D(X_2)$ and $\gamma_1>0$. E.g. if we choose $D(x)=1$, then
$B(x)\propto \exp(cx)$, if $D(x)=x$, then $B(x)\propto x^c$ with $c\neq0$, if
$D(x)=x^q$, $0\leq q<1$, then $B(x)\propto \exp(\frac c{1-q}x^{1-q})$. 
We focus on the most natural choice (i.e. $B$ not being of   exponential type) and therefore pick
\begin{equation*}
B(X_2)= a X_2^\alpha \quad \text{and} \quad D(X_2)=b\rho X_2.
\end{equation*}
% Note that the Heston model does not correspond to one of the three
% cases. Another possibility would be to assume instead that $A(X_1)$ is linear
% and then see for what coefficients all ODEs can be solved explicitly; the
% Heston model would then fall in this class. However, since we want to go
% beyond the case where $A(X_1)$ is linear, we proceed by considering the
% choices of $B(X_2)$ and $D(X_2)$ made above. 
%%%%%%%%%%%%%%%%%%%%5
These choices give us
\begin{equation*}
\begin{split}
\mathrm{d}X_1(t)= &    -\frac12 A(X_1) A'(X_1)a^2 X_2^{2\alpha} \mathrm{d}t 
+ A(X_1)a X_2^\alpha \circ \mathrm{d} \left( B^1_t -\frac12\alpha b\rho t
\right)  \\ 
\mathrm{d}X_2(t)= &  \left[ C(X_2)-\frac12 \left[ b^2\rho^2 X_2
    +E(X_2)E'(X_2) \right]  \right] \mathrm{d}t +  b\rho X_2 \circ
\mathrm{d}B_t^1 + E(X_2) \circ \mathrm{d}B_t^2. 
\end{split}
\end{equation*}
%%%%%%%%%%%%%%%%%%
With
\begin{equation*}
(V_0^1,V_0^2)= \left(-\frac12 A(x_1) A'(x_1)a^2 x_2^{2\alpha}    ,
  C(x_2)-\frac12 [ b^2\rho^2 x_2 +E(x_2)E'(x_2)] \right), 
\end{equation*}
define  $h(t;x_2)=x_2\mathrm{e}^{tV_0^2}$. We would like $\int_0^t
h(s;x_2)^{2\alpha}\mathrm{d}s$ to have an explicit expression, since it will
appear in the first component of $x\mathrm{e}^{tV_0}$. Possible cases are (i)
$h(t;x_2)\propto t+x_2$, (ii) $h(t;x_2) \propto \mathrm{e}^{ct}$ or (iii) very
specific cases like $h(t;x_2)=p(t,x_2)^{\frac1{2\alpha}}$ with $p$ nice. Both
(i) and (ii) lead to  $C$ being affine and $E$ being affine or of
square root type. We pick 
\begin{equation*}
C(X_2)=\kappa(\theta-X_2) \quad \text{and}   \quad E(X_2)=b\sqrt{1-\rho^2}X_2.
\end{equation*}
and we shall later motivate why we let $E$ be linear. With these choices
we can write 
\begin{equation*}
\begin{split}
\mathrm{d}X_1(t)= &    -\frac12 A(X_1) A'(X_1)a^2 X_2^{2\alpha} \mathrm{d}t 
+ A(X_1)a X_2^\alpha \circ \mathrm{d} \left( B^1_t -\frac12\alpha b\rho t
\right)  \\ 
\mathrm{d}X_2(t)= &  \left[\kappa(\theta-X_2) -\frac12  b^2  X_2   \right]
\mathrm{d}t +  b\rho X_2 \circ \mathrm{d}B_t^1 + b\sqrt{1-\rho^2} X_2
\circ \mathrm{d}B_t^2. 
\end{split}
\end{equation*}
We now rewrite the second SDE in the form 
\begin{equation*}
\dd X_2(t)=H(X_2)\mathrm{d}t + +  b\rho X_2 \circ \mathrm{d} \left( B^1_t
  +\gamma_1 t \right)  
+  b\sqrt{1-\rho^2} X_2  \circ \mathrm{d} \left(  B_t^2 + \gamma_2 t  \right)
\end{equation*}
with $\gamma_1=-\frac12\alpha b\rho $ (as in the first SDE) and with
$\gamma_2$ a constant such that $H(X_2)$ becomes as simple as possible (recall
that we want $\int_0^t (x_2\mathrm{e}^{sH})^{2\alpha}\mathrm{d}s$ to be
explicit). We get 
\begin{multline*}
 \mathrm{d}X_2(t)=  \kappa\theta  \mathrm{d}t +  b\rho X_2 \circ \mathrm{d}
 \left( B^1_t -\frac12\alpha b\rho t \right) \\  
+  b\sqrt{1-\rho^2} X_2  \circ \mathrm{d} \left(  B_t^2 + \frac{ \alpha
    b\rho^2 - 2\kappa/b - b}{2\sqrt{1-\rho^2}}  t  \right). 
\end{multline*}
Note that if we would have chosen $E$ to be affine but not linear or have
chosen $E$ of square root type, we would not be able to make $H$ so
simple. (For the same reason we have chosen $D$ linear and not generally
affine.)

Finally, $A(X_1)$ is left to choose. Since we want to end up with  a model accessible to SCFC, the function $A$ should be such that
$x_1\mathrm{e}^{tA}$ is explicit, which means that we want 
$\int_{\cdot}^x \frac{\mathrm{d}y}{A(y)}$ to have an explicit inverse. Also,
$x_1\mathrm{e}^{tA\cdot A'}$ should be explicit. The obvious candidates are
$A(X_1)=X_1^\beta$, $A(X_1)=\mathrm{e}^{cX_1}$ and $A(X_1)=X_1+c$ and all lead to models that are accessible to SCFC. %For our model, we haven chosen the first one. 
As a case study we choose the first one and apply the NV scheme with drift to the resulting model in the next section.

\subsection{Generalized SABR (with shifted log-normal 2nd
  factor)}\label{sv:gensabr} 
In the previous section we constructed a particular class of SV-models which are accessible to SCFC, namely: 
\begin{equation*}
\begin{split}
\mathrm{d}X_1(t)= &    a  X_2(t)^{\alpha} X_1(t)^{\beta} \mathrm{d}B^1_t  \\
\mathrm{d}X_2(t)= &  \kappa(\theta-X_2(t))\mathrm{d}t +  b X_2(t) \left(
  \rho\mathrm{d}B_t^1 + \sqrt{1-\rho^2} \mathrm{d}B_t^2 \right), 
\end{split}
\end{equation*}
with $X_1(0)=x_1$ and $X_2(0)=x_2$. We assume that the parameters satisfy
$\frac12\leq \beta\leq1$, $\theta, \kappa\geq0$, $\alpha>0$, $a,b>0$,
$-1<\rho<1$. 
%Note that the SDE for $X_2$ is the so-called Brennan-Schwartz SDE. 
Details
surrounding well-posedness, integrability properties and martingale properties
can be found in Lions and Musiela
\citelist{\cite{lionsmusiela}\cite{lionsmusiela2}}. 
A simple application of It\^o's formula shows that
\begin{equation*}
X_2(t)=
 x_2\mathrm{e}^{-(\kappa+\frac12 b^2)t + b W_t}  + \kappa\theta\int_0^t
 \mathrm{e}^{-(\kappa+\frac12 b^2)(t-s)}\mathrm{e}^{b(W_t- W_s)}
 \mathrm{d}s, 
\end{equation*}
where $W_t= \rho B_t^1 + \sqrt{1-\rho^2} B_t^2$ and thus $X_2(t)>0$ for all
$t\geq0$, provided $x_2>0$.  

We shall now give all the  ODE solutions that are required to apply the NV scheme (with drift).
First note that the vector fields $V_1$ and $V_2$ corresponding to $B^1$ and $B^2$ are given by
\begin{equation*} 
V_1=\begin{pmatrix}
 a  x_2^{\alpha} x_1^{\beta}  \\
 b  \rho  x_2 
\end{pmatrix}, \quad
 V_2=\begin{pmatrix}
 0 \\
 b \sqrt{1-\rho^2}  x_2  
\end{pmatrix}
\end{equation*}
and the ODE solutions are
\begin{equation*}
\mathrm{e}^{sV_1}x=
\begin{pmatrix}
 g_1(s) \\
 x_2\exp \left( b  \rho  s \right) 
\end{pmatrix},
\end{equation*}
\begin{equation*}
\mathrm{e}^{sV_2}x=
\begin{pmatrix}
x_1 \\
 x_2\exp \left( b \sqrt{1-\rho^2} s \right)  
\end{pmatrix},
\end{equation*}
where
\begin{equation*}
\begin{split}
 g_1(s) = & \left[ (1-\beta)\frac{a x_2^\alpha}{\alpha b\rho} \left(
     \mathrm{e}^{\alpha b\rho s}-1 \right) +x_1^{1-\beta}
 \right]^{1/(1-\beta)}_+, \quad \frac12\leq \beta<1, \\ 
 g_1(s) = & x_1\exp \left( \frac{a x_2^\alpha}{\alpha b\rho} \left(
     \mathrm{e}^{\alpha b\rho s}-1 \right)  \right)  , \quad \beta=1. 
\end{split}
\end{equation*}
The It\^o drift vector field $\widetilde{V}_0$ and Stratonovich
drift vector field $V_0$  of $X$ are given by 
\begin{equation*}
\widetilde{V}_0(x) =
\begin{pmatrix}
0 \\
\kappa\theta- \kappa x_2 \\
\end{pmatrix}, \quad
V_0(x) =
\begin{pmatrix}
 -\frac12 a^2\beta  x_2^{2\alpha} x_1^{2\beta-1} -\frac12\alpha ab\rho
 x_2^\alpha x_1^\beta \\ 
\kappa\theta-(\kappa+\frac12 b^2)x_2 
\end{pmatrix}.
\end{equation*}
We have $\mathrm{e}^{sV_0}x=(H(s), h(s))^T$ with
\begin{equation*}
\begin{split}
h(s)= & \left( x_2-\frac{\kappa\theta}{\kappa+\frac12 b^2} \right)
\mathrm{e}^{-(\kappa+\frac12 b^2)s}+\frac{\kappa\theta}{\kappa+\frac12
  b^2} 
\end{split}
\end{equation*}
and $H$ needs to be numerically solved. (We have already pointed to this
difficulty when we discussed the classical SABR example earlier on.) 

\bigskip

Let us now show that by using Brownian increments with drift, this problem can
be resolved: all necessary flows can be computed in closed form. 
Recall from the previous section that we can rewrite $X$ as  
\begin{equation*}
\begin{split}
\mathrm{d}X_1(t)= &    -\frac12 a^2 \beta X_2^{2\alpha}X_1^{2\beta-1} \mathrm{d}t 
+  a X_2^\alpha X_1^\beta \circ \mathrm{d} \left( B^1_t + \gamma_1 t \right)  \\
\mathrm{d}X_2(t)= & \kappa\theta  \mathrm{d}t +  b\rho X_2 \circ \mathrm{d}
\left( B^1_t +\gamma_1 t \right) 
 +  b\sqrt{1-\rho^2} X_2  \circ \mathrm{d} \left(  B_t^2 + \gamma_2  t
 \right). 
\end{split}
\end{equation*}
with
\begin{equation*}
\gamma_1=-\frac12\alpha b\rho \quad \text{and} \quad \gamma_2= \frac{ \alpha
  b\rho^2 - 2\kappa/b - b}{2\sqrt{1-\rho^2}}. 
\end{equation*}
We see here that the assumption $-1<\rho<1$ is crucial.
Note that the vector fields corresponding to $B^1_t+\gamma_1 t$ and
$B^2_t+\gamma_2 t$, respectively, are $V_1$ and $V_2$. 
Denote by $V_0^{(\gamma)}$ the remaining part, i.e.
\begin{equation*}
V_0^{(\gamma)}(x) =
\begin{pmatrix}
 -\frac12 a^2\beta  x_2^{2\alpha} x_1^{2\beta-1} \\
\kappa\theta  
\end{pmatrix}.
\end{equation*}
Then we have $\mathrm{e}^{sV_0^{(\gamma)}}x=(g_0 ,\kappa\theta s+x_2)^T$ with (cf. Appendix)
\begin{equation*}
\begin{split}
 g_0(s) = & \left(   {-a^2\beta(1 -\beta )}   P(s)  + x_1^{2(1-\beta)}
 \right)^{\frac1{2(1-\beta)}}_+, \quad \frac12< \beta<1, \\ 
 g_0(s) = & x_1 \exp  \left(  -\frac12 a^2    P(s) \right), \quad \beta=1, \\
g_0(s)= &   {-\frac14 a^2}   P(s)  + x_1 , \quad \beta=\frac12,
\end{split}
\end{equation*}
where
\begin{equation*}
P(s)= \frac1{(2\alpha+1)\kappa\theta}\left( (\kappa\theta
  s+x_2)^{2\alpha+1}-x_2^{2\alpha+1} \right). 
\end{equation*}
Note that when $\kappa=0$ or $\theta=0$, $P(s)$ should be understood in the
limiting sense, i.e., $P(s) = s x_2^{2\alpha}$ for $\kappa = 0$ or $\theta =
0$.

\begin{remark}
  Since the SABR model is a special case of the model presented here --
  corresponding to  $\alpha = 1$, $\kappa = 0$ -- the semi-closed
  form NV algorithm developed above can be, in particular, applied to the
  SABR model.
\end{remark}

\subsection{Girsanov transform}

We have seen for the example above, that if one uses the standard NV scheme,
the flow of the drift vector field is not available in closed form. Besides
using the `drift trick' as we did above, it is also possible to absorb this
drift in a change-of-measure; the details of this are outlined below. There
is, however, a serious downside to this: the Girsanov-density which appears
due to the change-of-measure will \emph{add significantly to the variance of the
object to be sampled}. Thus, without further variance reduction, we do not
advertise the use of the Girsanov transform in this context.

Let $Y=(Y_1,Y_2)$ be the process defined by 
\begin{equation*}
\begin{split}
\mathrm{d}Y_1(t)= &    a  Y_2(t,x)^{\alpha} Y_1(t)^{\beta} \left( -\gamma_1
  \mathrm{d}t +  \mathrm{d}B^1_t \right)   \\ 
\mathrm{d}Y_2(t)= &      \kappa(\theta-Y_2(t))\mathrm{d}t +b \rho Y_2(t)
\left( -\gamma_1 \mathrm{d}t +  \mathrm{d}B^1_t \right) +
b\sqrt{1-\rho^2}Y_2(t)  \left( -\gamma_2 \mathrm{d}t +  \mathrm{d}B^2_t
\right). 
\end{split}
\end{equation*}
and let $\mathbb{P}$ be the probability measure under which $B=(B^1,B^2)$ is a
2-dimensional standard Brownian motion. Define the  probability measure
$\mathbb{Q}$ by  
$\frac{\mathrm{d}\mathbb{Q}}{\mathrm{d}\mathbb P}|_{\mathcal{F}_t}=\mathcal
E(t)$, where 
\begin{equation*}
\mathcal{E}(t)=
\exp \left(  \gamma_1 B_t^1 + \gamma_2 B_t^2 - \frac12
  (\gamma_1^2+\gamma_2^2)t \right). 
\end{equation*}
Then by Girsanov, under $\mathbb{Q}$, $(B^1,B^2)$ is equal in law to a
2-dimensional standard Brownian motion plus constant drift equal to
$(\gamma_1,\gamma_2)$.  
Hence under $\mathbb{Q}$, $Y$ is equal in law to $X$ under $\mathbb{P}$ and we
have for $f:{\mathbb{R}^2}\rightarrow\mathbb{R}$ measurable, 
\begin{equation*}
\mathbb{E}^{\mathbb{P}}[f(X(t))]=\mathbb{E}^{\mathbb{Q}}[f(Y_1(t),Y_2(t))] =
\mathbb{E}^{\mathbb{P}}[f(Y_1(t),Y_2(t))\mathcal{E}(t)]. 
\end{equation*}
Hence a ``weighted'' NV scheme with explicit solutions to all ODEs can be
obtained by using the NV scheme for the process $Y$ and then multiplying the
payoff $f(Y_1(t),Y_2(t))$, as is done in importance sampling, by $\mathcal E(t)$. Note that all the ODE solutions corresponding to
the NV scheme for $Y$ are explicit, since the vector fields corresponding to
$Y$ are $V_0^{(\gamma)}$, $V_1$ and $V_2$. 

To back up our claim about the additional variance caused by the Girsanov
density $\mathcal{E}(t)$, note that $\mathbb{V}(\mathcal{E}(t)) =
e^{(\gamma_1^2 + \gamma_2^2) t} - 1$, which is only negligible when $\gamma_1$
and $\gamma_2$ are close to zero.

\subsection{A multi-dimensional version}\label{sv:gensabrmd}

Let us illustrate how the techniques introduced (until now in the
context of 2-dimensional models) remain feasible in typical higher dimensional models (what we have in mind here is some multi asset SV model). Since
it is precisely the curse of dimensionality that forces one to use stochastic
methods (rather than PDE methods, say) we want to be fully explicit in showing
how our ideas are implemented in higher dimensions. More specifically, we
shall consider the following multi-dimensional version of our SABR-type model:
for $i=1,\ldots,N$, 
\begin{equation*}
\begin{split}
\mathrm{d}X_i(t)= &    a_i  Y_i(t)^{\alpha_i} X_i(t)^{\beta_i}
\mathrm{d}\widetilde{B}^i_t  \\ 
\mathrm{d}Y_i(t)= &  \kappa_i(\theta_i-Y_i(t) )\mathrm{d}t +  b_i Y_i(t)
\mathrm{d}\widetilde{W}_t^i,
\end{split}
\end{equation*} 
with $X_i(0)=x_i$, $Y_i(0)=y_i$ and $\frac12\leq \beta_i\leq1$, $\theta_i,\kappa_i\geq0$, $\alpha_i>0$, $a_i,b_i>0$,
$-1<\rho_i<1$. Here $(\begin{smallmatrix} \widetilde{\mathbf{B}} \\
  \widetilde{\mathbf{W}} \end{smallmatrix})$, with $
\widetilde{\mathbf{B}}=(\widetilde{B}^1,\ldots,\widetilde{B}^N)^T$ and $
\widetilde{\mathbf{W}}=(\widetilde{W}^1,\ldots,\widetilde{W}^N)^T$ is a
2N-dimensional Brownian motion with correlation matrix given by ${\rho}$  which we assume to be positive-definite. Let
$\sqrt{\rho}$ be the unique lower-triangular matrix such that
$\sqrt{\rho}\sqrt{\rho}^T=\rho$ (Choleski decomposition). Then
$(\begin{smallmatrix} \widetilde{\mathbf{B}} \\
  \widetilde{\mathbf{W}} \end{smallmatrix})\stackrel{(D)}{=} \sqrt{\rho}
(\begin{smallmatrix} {\mathbf{B}} \\ {\mathbf{W}} \end{smallmatrix})$, where
$(\begin{smallmatrix} {\mathbf{B}} \\ {\mathbf{W}} \end{smallmatrix})$, with $
{\mathbf{B}}=({B}^1,\ldots,{B}^N)^T$ and $
{\mathbf{W}}=({W}^1,\ldots,{W}^N)^T$ is a $2N$-dimensional standard Brownian
motion. Hence we can write for $i=1,\ldots,N$, 
\begin{equation*}
\begin{split}
\mathrm{d}X_i(t)= &    a_i  Y_i(t)^{\alpha_i} X_i(t)^{\beta_i} \mathrm{d} \left(
  \sum_{j=1}^N \sqrt{\rho}_{i,j}{B}^j_t  + \sum_{j=1}^N \sqrt{\rho}_{i,N+j}
  W_t^j \right)  \\ 
\mathrm{d}Y_i(t)= &  \kappa_i(\theta_i-Y_i(t))\mathrm{d}t +  b_i Y_i(t)
\mathrm{d} \left( \sum_{j=1}^N \sqrt{\rho}_{N+i,j}{B}^j_t  + \sum_{j=1}^N
  \sqrt{\rho}_{N+i,N+j} W_t^j \right). 
\end{split}
\end{equation*}
Let for $j=1,\ldots,N$, $V_j$ and $U_j$ be the vector fields corresponding to
$B^j$ and $W^j$, respectively. We have 
\begin{equation*}
V_j(x,y) =
\begin{pmatrix}
  a_1  y_1^{\alpha_1} x_1^{\beta_1} \sqrt{\rho}_{1,j}  \\
 b_1  y_1 \sqrt{\rho}_{N+1,j}  \\
\vdots \\
a_N  y_N^{\alpha_N} x_N^{\beta_N} \sqrt{\rho}_{N,j}  \\
 b_N  y_N \sqrt{\rho}_{2N,j}  
\end{pmatrix}
, \quad 
U_j(x,y)=
\begin{pmatrix}
  0  \\
 b_1  y_1 \sqrt{\rho}_{N+1,N+j}  \\
\vdots \\
0  \\
 b_N  y_N \sqrt{\rho}_{2N,N+j}  
\end{pmatrix}
.
\end{equation*}
It follows that
\begin{equation*}
V_0(x,y)=
\begin{pmatrix}
 -\frac12 p_1 a_1^2\beta_1  y_1^{2\alpha_1} x_1^{2\beta_1-1}   - \frac12
 q_1\alpha_1 a_1b_1 y_1^{\alpha_1} x_1^{\beta_1} \\ 
\kappa_1\theta_1-(\kappa_1 +\frac12 b_1^2 r_1)y_1  \\
\vdots \\
-\frac12 p_N a_N^2\beta_N  y_N^{2\alpha_N} x_N^{2\beta_N-1}   - \frac12
q_N\alpha_N a_Nb_N y_N^{\alpha_N} x_N^{\beta_N} \\ 
\kappa_N\theta_N-(\kappa_N +\frac12 b_N^2 r_N)y_N  
\end{pmatrix}
,
\end{equation*}
where for $i=1,\ldots,N$,
\begin{equation*}
\begin{split}
p_i= & \sum_{j=1}^N \left[ \left( \sqrt{\rho}_{i,j} \right)^2 + \left(
    \sqrt{\rho}_{i,N+j} \right)^2 \right] = \rho_{i,i}=1, \\ 
q_i = & \sum_{j=1}^N \left[ \sqrt{\rho}_{N+i,j}\sqrt{\rho}_{i,j} +
  \sqrt{\rho}_{N+i,N+j}\sqrt{\rho}_{i,N+j} \right] = \sum_{j=1}^N  \sqrt{\rho}_{N+i,j}\sqrt{\rho}_{i,j}, \\ 
r_i= & \sum_{j=1}^N  \left[ \left( \sqrt{\rho}_{N+i,j} \right)^2 + \left(
    \sqrt{\rho}_{N+i,N+j} \right)^2 \right] =\rho_{N+i,N+i}=1. 
\end{split}
\end{equation*}

We would like to write the system $(X_i,Y_i)$, $i=1,\ldots,N$ in the following
way 
\begin{equation*}
\begin{split}
\mathrm{d}X_i(t)= &    -\frac12  a_i^2 \beta_i
Y_i(t)^{2\alpha_i} X_i(t)^{2\beta_i-1} \mathrm{d}t +  a_i Y_i(t)^{\alpha_i} X_i(t)^{\beta_i} \\
& \times \left(  \sum_{j=1}^N \sqrt{\rho}_{i,j}
  \circ \mathrm{d}  \left\{ B^j_t +\gamma_j t \right\}  + \sum_{j=1}^N
  \sqrt{\rho}_{i,N+j}  \circ \mathrm{d}  \left\{ W_t^j + \delta_j t \right\}
\right)  \\ 
\mathrm{d}Y_i(t)= & \kappa_i\theta_i  \mathrm{d}t +  b_i Y_i(t) \\
& \times \left(
  \sum_{j=1}^N \sqrt{\rho}_{N+i,j} \circ \mathrm{d}   \left\{ B^j_t +\gamma_j
    t \right\}  + \sum_{j=1}^N \sqrt{\rho}_{N+i,N+j} \circ \mathrm{d} \left\{
    W_t^j + \delta_j t \right\} \right), 
\end{split}
\end{equation*}
for a certain $\vec\gamma=(\gamma_1,\ldots,\gamma_N)$ and $\vec\delta=(\delta_1,\ldots,\delta_N)$.
Looking at $V_0$, we should choose $\vec{\gamma}$ and $\vec{\delta}$ such that
for $i=1,\ldots,N$, 
\begin{equation*}
 \begin{split}
 \sum_{j=1}^N \sqrt{\rho}_{i,j}\gamma_j + \sum_{j=1}^N \sqrt{\rho}_{i,N+j}
 \delta_j = & -\frac12 q_i\alpha_i b_i, \\  
\sum_{j=1}^N \sqrt{\rho}_{N+i,j}\gamma_j + \sum_{j=1}^N \sqrt{\rho}_{N+i,N+j}
\delta_j = & -\frac{\kappa_i+\frac12 b_i^2 }{b_i}. 
 \end{split}
\end{equation*}
These are $2N$ linear equations with $2N$ unknowns. It follows that there
exists a unique $\vec{\gamma}$ and $\vec{\delta}$ such that the above
equalities are satisfied if $\sqrt{\rho}$ is of full rank. This is the
multi-dimensional analogue of the condition $-1<\rho<1$ of the previous
section. We see that in order to apply the NV scheme with drift, we need to find the flows corresponding to the vector fields $V_j(x,y)$, $U_j(x,y)$ and
\begin{equation*}
V_0^{(\vec \gamma,\vec \delta)}(x,y)=
\begin{pmatrix}
 -\frac12  a_1^2\beta_1  y_1^{2\alpha_1} x_1^{2\beta_1-1}    \\ 
\kappa_1\theta_1  \\
\vdots \\
-\frac12  a_N^2\beta_N  y_N^{2\alpha_N} x_N^{2\beta_N-1}   \\ 
\kappa_N\theta_N  
\end{pmatrix}.
\end{equation*}
All the solutions to these ODEs can be found explicitly as in Section \ref{sv:gensabr}.

\subsection{Numerical analysis of our NV scheme with drift}
\label{sec:numer-analys-drift}

In this section 
 we want to prove second order weak
convergence of $\mathbb E\left( f\left(\overline X^{(NVd),K}(T,x) \right) \right)$ as $K \to
\infty$ for smooth $f$ with $\overline X^{(NVd),K}(T,x)$ given by \eqref{eq:NV-drift}-\eqref{eq:NV-driftpart2}. As in the original proof by Ninomiya and Victoir \cite{ninomiyavictoir} for the classical NV scheme, we use a Taylor
expansion to get the local order of the weak error by comparison with the
known local weak order of the classical Ninomiya-Victoir scheme. Let
$\overline X^{(NV),K}(T/K,x)$ be as in Section \ref{sectionNVscheme}. Then the difference in the Taylor expansion of the
expectation in one step is given by
\begin{multline}
  \label{eq:weak-order-drift}
  \mathbb E\left(\left.f\left(\overline X^{(NVd),K}(T/K,x) \right) \right| \Lambda_1 = -1 
  \right) - 
  \mathbb E\left(\left.f\left(\overline X^{(NV),K}(T/K,x) \right) \right| \Lambda_1 = -1 
  \right) = \\
  \Biggl[\f{1}{2} \sum_{1 \le i < j \le d} \gamma_i \gamma_j V_iV_j f(x)-
    \f{1}{2} \sum_{1 \le j < i \le d} \gamma_i \gamma_j V_iV_j f(x) +\\
    + \f{1}{4} \sum_{1 \le i < j \le d} \gamma_i V_i (V_j)^2 f(x) -
    \f{1}{4} \sum_{1 \le j < i \le d} \gamma_i V_i (V_j)^2 f(x) +\\
    + \f{1}{4} \sum_{1 \le i < j \le d} \gamma_j (V_i)^2V_j f(x) -
    \f{1}{4} \sum_{1 \le j < i \le d} \gamma_j (V_i)^2V_j f(x) 
  \Biggr] \left(T/K \right)^2 + \mathcal{O}\left(\left(T/K\right)^3
  \right).
\end{multline}
When we condition on $\Lambda_1 = 1$, only the order of the indices are
swapped. By the peculiar structure of~\eqref{eq:weak-order-drift}, this means
that the signs of all terms of the difference change, i.e.,
\begin{multline*}
   \mathbb E\left(\left.f\left(\overline X^{(NVd),K}(T/K,x) \right) \right| \Lambda_1 = 1 
  \right) - 
  \mathbb E\left(\left.f\left(\overline X^{(NV),K}(T/K,x) \right) \right| \Lambda_1 = 1 
  \right) =\\
  -\left[ \mathbb E\left(\left.f\left(\overline X^{(NVd),K}(T/K,x) \right) \right| \Lambda_1 = -1 \right) - 
  \mathbb E\left(\left.f\left(\overline X^{(NV),K}(T/K,x) \right) \right| \Lambda_1 = -1, 
  \right) \right] \\ + \mathcal{O}\left(\left(T/K\right)^3\right).
\end{multline*}
Thus, taking the unconditional expectation in $\Lambda_1$ gives
\begin{equation*}
  \label{eq:weak-order-drift-2}
  \mathbb E\left( f\left(\overline X^{(NVd),K}(T/K,x) \right)  
  \right) - 
  \mathbb E\left( f\left(\overline X^{(NV),K}(T/K,x) \right)  
  \right) = \mathcal{O}\left(\left(T/K\right)^3
  \right),
\end{equation*}
and second order convergence of the Ninomiya-Victoir scheme with drift follows
exactly as in the case without drift.

\begin{remark}
  \label{rem:cbature-with-drift}
  The drift trick also works for classical cubature on Wiener space as in
  Lyons-Victoir~\cite{lyo/vic04}, i.e., when $W^0(t) \equiv t $, simply by
  adding the same drift $t \gamma_i$ to the $i^{\text{th}}$ component of the
  cubature path $W$. This procedure retains the original order of convergence
  for the given cubature formula, as can be trivially seen by comparing the
  ODEs with and without drift. Note that in the Ninomiya-Victoir scheme a
  slightly more difficult argument as discussed above is necessary, since here
  the ``time'' component $W^0$ is not linear.
\end{remark}

\section{Numerical results}

In this section we report the results of our numerical experiments. For this
we have chosen three models: the SABR model, the generalized SABR model and
the multi-dimensional generalized SABR model.  The numerical results for these
models are given in Section \ref{numres:sabr}, Section \ref{numres:gensabr}
and Section \ref{numres:gensabrmd}, respectively. For each of the models,
we compare the NV scheme with drift to the regular NV scheme and the Euler
scheme and in all the experiments reported in this paper, we used Quasi
Monte-Carlo for the integration. In order to check the order of the three
schemes, we first give plots of the relative discretization error against the
number of time steps $K$.  For these plots the number of `simulated
trajectories' $M$ is chosen such that the integration error is negligible
compared to the discretization error. (Since we are using Quasi Monte Carlo,
$M$ is strictly speaking not the number of simulations, but the size of the
finite low-discrepancy sequence used for the computation.) In order to compute
these errors we of course need to know the `true value' (or a good estimate of
it).  For the two experiments involving the generalized SABR model, the true
value was obtained by running the code longer than reported in the
plots. However, in the case of the SABR model, the true value was estimated by
extrapolation of the values obtained by the code (the SABR formula is not
exact enough!)

Results comparing the discretization error against the number of time steps,
might not really seem practically relevant, since they only relate the number
of time-steps for different numerical methods, but not the corresponding
computational cost or computer time. Therefore we also give tables of the
computational time of the different schemes. If we want to compare the
run-time for different methods, we need to ensure the fairness of the
comparison, i.e., we need to compare the different methods with parameters
giving similar computational errors.  We basically have two parameters for the
numerical method, namely the number of time-steps $K$ and the number of
simulated trajectories $M$ -- where every trajectory is given by a
$D(K)$-dimensional vector from a low discrepancy sequence, in our case the
Sobol-sequence. (Here, $D(K)$ depends on both the model and the method.)
Unlike for Monte Carlo simulation, where accurate, but probabilistic error
estimates are available, there is no simple error estimation procedure for QMC
(as far as we are aware). Therefore, it is not obvious how to choose the
number of trajectories for a comparison of run-times. Choosing the same number
of trajectories for every method might not be appropriate, because some
methods might yield ``rougher'' integration problems, requiring a higher
number of trajectories for a comparable precision.

In our comparison we proceeded as follows. For a given model and method we
first choose $K$ (and $M$ very large) such that the relative discretization error is around $10^{-3}$. For
simplicity, we take the computations run for producing the plots, which means
that $K$ is a power of two. Then we start with $M_0 = 1000$ trajectories, run
the method, and double the number $M$ of trajectories until the observed
(absolute) error is \emph{consistently} closer than $2 \times 10^{-5}$ to the
true (absolute) error for the fixed given $K$. 

To explain this in more detail, first recall that the computational error in (Quasi) Monte Carlo simulations for
SDEs splits into two parts: $\text{Error} = \text{Error}_{\text{dis}} +
\text{Error}_{\text{int}}$. Here, $\text{Error}_{\text{dis}}$ stems from the
time-discretization of the SDE. This part of the error is controlled by
$K$. $\text{Error}_{\text{int}}$ is the error from the integration, i.e., from
the numerical computation of the expectation of the solution to the
discretized SDE. This error part is controlled by $M$. In the comparison
procedure, we first fix $K$ such that the relative discretization error
$\text{Error}_{\text{dis}} / C$ ($C$ denoting the true result) is around
$10^{-3}$. Then we choose $M$ (by a doubling procedure) such that
$\text{Error}_{\text{int}} \le 2 \times 10^{-5}$. (Since in all cases $C
\approx 0.1$, this means we choose the integration error to be one fifth of
the discretization error.) 

In the following tables, we report the found parameters, the corresponding
relative error and the computational time in seconds -- all computations where
performed on the same computer, a Toshiba laptop with 6 GB RAM and four Intel
Core i7 CPUs with 1.6 GHz. Notice that the run-time scales linearly with the
number $M$ of trajectories.

\subsection{SABR}\label{numres:sabr}
In this section we give the results corresponding to the experiment with the
SABR model. Recall that this SV-model is given by 
\begin{equation*}
\begin{split}
\mathrm{d}X_1(t)= &    a  X_2(t) X_1(t)^{\beta} \mathrm{d}B^1_t  \\
\mathrm{d}X_2(t)= &    b X_2(t) \left( \rho\mathrm{d}B_t^1 + \sqrt{1-\rho^2}
  \mathrm{d}B_t^2 \right), 
\end{split}
\end{equation*}
with $X_1(0)=x_1$ and $X_2(0)=x_2$ and where the parameters satisfy
$\frac12\leq \beta\leq1$, $a,b>0$, $-1<\rho<1$. The parameters chosen for
the experiment are $\beta=0.9$, $a=1.0$, $b=0.4$, $\rho=-0.7$, $x_1=1.0$,
$x_2=0.3$. As the derivative we choose a (European) call option with maturity
time $T=1.0$ and strike price $K=1.05$. For simplicity we assume that the
interest rate is zero. The corresponding estimated `true result' is
0.09400046. 

\begin{figure}[!htb] 
\centering  
\includegraphics[scale=.5]{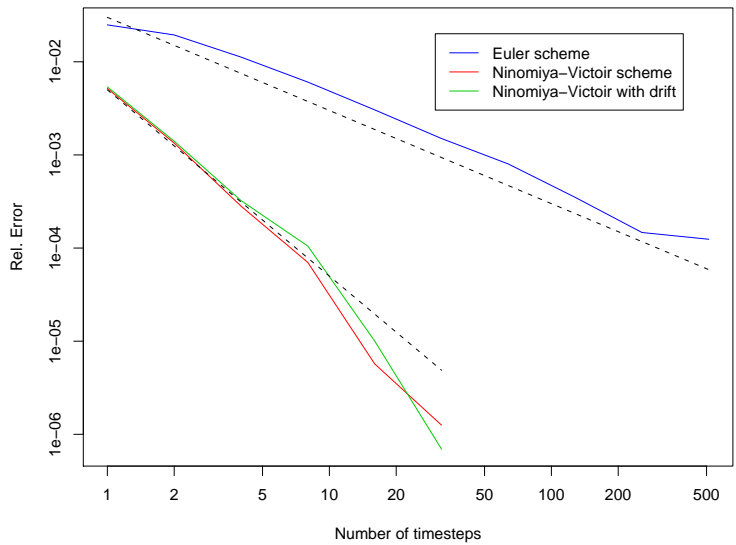}
\caption{Order of convergence for the SABR model.} 
\label{fig:sabr} 
\end{figure}

In Figure \ref{fig:sabr} the convergence rates of the three schemes is
graphically displayed. We clearly see the second-order convergence of the two
NV schemes compared to the first-order convergence of the Euler method.

\begin{table}[!htb]
  \centering
  \begin{tabular}{|l|c|c|c|c|}
    \hline
    Method & $K$ & $M$ & Rel.~Error & Time \\
    \hline
    Euler & $32$ & $512000$ & $0.00150$ & $5.87$ sec\\
    \hline
    Ninomiya-Victoir & $2$ & $512000$ & $0.00134$ & $2.44$ sec\\
    \hline
    NV with drift & $2$ & $128000$ & $0.00140$ & $0.28$ sec\\
    \hline
  \end{tabular}
  \caption{Computational time for the SABR model}
  \label{tab:sabr-time}
\end{table}
In Table~\ref{tab:sabr-time} the timings are reported for the SABR
model. Notice that the Ninomiya-Victoir method both in its original form and
in its variant are clearly more efficient than the Euler method, by a factor
two or three. On the other hand, the simpler structure of the Ninomiya-Victoir
method with drift results in a considerable speed up if compared with the
original method. Notice that a speed up with a factor two would even hold if
we reject the empirically found choice of $M$ for the drift variant and use
the same number of trajectories as for the other two methods.

\subsection{Generalized SABR} \label{numres:gensabr}

In this section we give the results corresponding to the experiment with the
generalized SABR model of Section \ref{sv:gensabr}. For convenience we restate
this SV-model: 
\begin{equation*}
\begin{split}
\mathrm{d}X_1(t)= &    a  X_2(t)^{\alpha} X_1(t)^{\beta} \mathrm{d}B^1_t  \\
\mathrm{d}X_2(t)= &  \kappa(\theta-X_2(t))\mathrm{d}t +  b X_2(t) \left(
  \rho\mathrm{d}B_t^1 + \sqrt{1-\rho^2} \mathrm{d}B_t^2 \right), 
\end{split}
\end{equation*}
with $X_1(0)=x_1$ and $X_2(0)=x_2$ and $\frac12\leq \beta\leq1$, $\theta>0$,
$\kappa\geq0$, $\alpha>0$, $a,b>0$, $-1<\rho<1$. For our experiment, we
choose the parameters as follows: $\beta=1.0$, $\theta=0.3$, $\kappa=2.0$,
$\alpha=0.5$, $a=1.0$, $b=0.5$, $\rho=-0.7$, $x_1=1.0$ and $x_2=0.2$.  We
further pick the same call option as in Section \ref{numres:sabr}.  The
estimated `true result' is 0.1767505855.

\begin{figure}[!htb]
\centering  
\includegraphics[scale=.5]{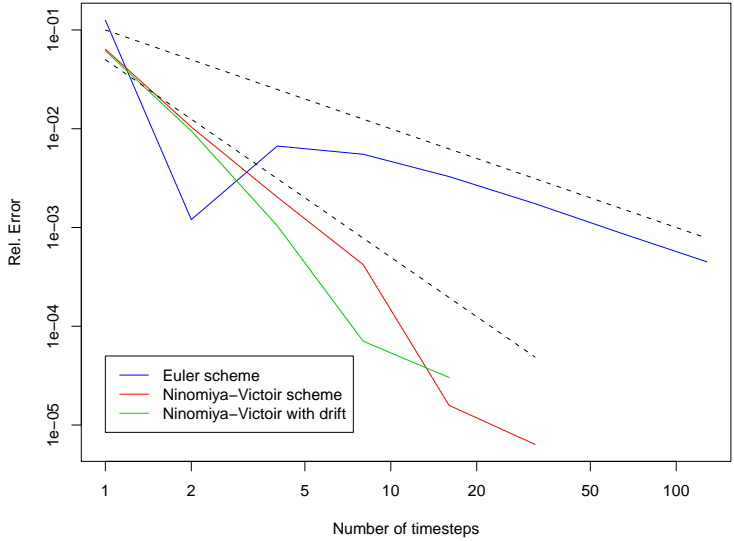}
\caption{Order of convergence for the generalized SABR model.} 
\label{fig:gensabr} 
\end{figure}

In Figure \ref{fig:gensabr} the discretization error against the number of
time steps is plotted. Again, we see the second-order convergence of the two
NV schemes compared to the first-order convergence of the Euler method.

\begin{table}[!htb]
  \centering
  \begin{tabular}{|l|c|c|c|c|}
    \hline
    Method & $K$ & $M$ & Rel.~Error & Time \\
    \hline
    Euler & $32$ & $8192000$ & $0.00174$ & $91.94$ sec\\
    \hline
    Ninomiya-Victoir & $4$ & $2048000$ & $0.00204$ & $13.93$ sec\\
    \hline
    NV with drift & $4$ & $1024000$ & $0.00104$ & $2.88$ sec\\
    \hline
  \end{tabular}
  \caption{Computational time for the generalized SABR model}
  \label{tab:gensabr-time}
\end{table}
In Table \ref{tab:gensabr-time} the timings are reported for the generalized
SABR model.  In the one-dimensional generalized SABR model, the
Ninomiya-Victoir method is only faster than the Euler method because the
integrand seems to be smoother. If one rejects our way to determine the
necessary number of trajectories $M$ as too crude and insists on taking the
same number for both methods, the Ninomiya-Victoir method will require almost
the same time as the Euler method in order to give comparable results at this
level. However, the Ninomiya-Victoir method with drift still retains a
convincing speed-up, again probably due to the simpler structure of the
subroutines.

\subsection{Multi-dimensional generalized SABR} \label{numres:gensabrmd}

In this section we give the numerical results corresponding to the
multi-dimensional generalized SABR model. 
Recall that from Section \ref{sv:gensabrmd} this model is given by
\begin{equation*}
\begin{split}
\mathrm{d}X_i(t)= &    a_i  Y_i ^{\alpha_i} X_i ^{\beta_i}
\mathrm{d}\widetilde{B}^i_t  \\ 
\mathrm{d}Y_i(t)= &  \kappa_i(\theta_i-Y_i )\mathrm{d}t +  b_i Y_i
\mathrm{d}\widetilde{W}_t^i, 
\end{split}
\end{equation*} 
$i=1,\ldots,N$. For our experiment we choose $N=4$ and as the derivative we
take a basket option with the same weight on each stock. % We refrain from
% quoting all the parameters due to the amount of them involved. In particular,
% we do not want to spell out the $8\times 8$ correlation matrix, which was
% randomly generated among positively definite correlation matrices but in such
% a way as to make sure that $\widetilde{B}^i$ and $\widetilde{W}^i$ are
% negatively correlated, as usual in equity modeling. The estimated `true value'
% of the basket option is $0.09254183$.
The parameters for the experiment have been chosen as
follows: $a = (1, 0.5, 0.3, 0.7)^T$, $b = (0.5, 0.8, 0.4, 0.6)^T$, $\alpha =
(0.5, 1, 0.7, 0.8)^T$, $\beta = (0.6, 0.7, 0.8, 0.9)^T$, $\kappa = (0.2, 0.7,
0.5, 0.9)^T$, $\theta = (0.3, 0.4, 0.6, 0.2)^T$, and, finally,
\begin{equation*}
  \rho \approx
  \begin{pmatrix}
    1 & 0.0111 & 0.6395 & -0.1081 & -0.3414 & -0.0642 & -0.2054 & -0.0236 \\
    0.0111 & 1 &  0.2698 & 0.2770 & 0.1651 & -0.3504 & -0.8186 & -0.4383 \\
    0.6395 & 0.2698 & 1 & -0.1381 & -0.1379 & -0.0031 & -0.3169 & -0.0161 \\
    -0.1081 & 0.2770 & -0.1381 & 1 & 0.7312 & -0.9030 & 0.0419 & -0.8121 \\
    -0.3414 &  0.1651 & -0.1379 & 0.7312 & 1 & -0.5969 &  0.0747 & -0.6703 \\
    -0.6420 & -0.3504 & -0.0031 & -0.9030 & -0.5969 & 1 & 0.1878 & 0.8790 \\
    -0.2054 & -0.8186 & -0.3169 & 0.0419 & 0.0747 & 0.1878 & 1 & 0.2796 \\
    -0.0236 & -0.4383 & -0.0161 & -0.8121 & -0.6703 & 0.8790 & 0.2796 & 1\\
  \end{pmatrix}.
\end{equation*}
Note that the above choice of $\rho$ implies that $\widetilde{B}^i$ and
$\widetilde{W}^i$ are negatively correlated, as usual in equity
modeling. Moreover, $\rho$ is positive definite -- and in fact, chosen at
random among all such correlation matrices. The estimated `true value'
of the basket option is $0.09254183$.

\begin{figure} 
\centering  
\includegraphics[scale=.5]{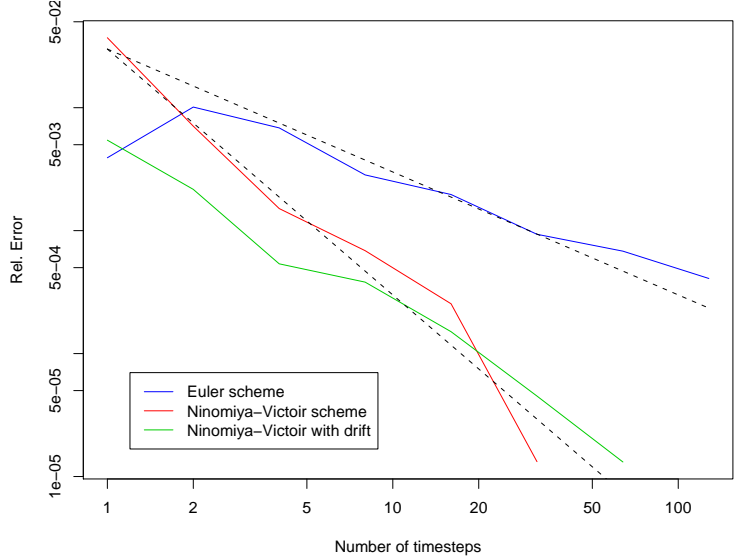}
\caption{Order of convergence for the multi-dimensional generalized SABR model.} 
\label{fig:gensabrmd} 
\end{figure}

The convergence rates of the three different schemes for this experiment are
graphically displayed in Figure \ref{fig:gensabrmd}. The picture is similar as
in the two previous cases in the sense that there is second-order convergence
for the two NV schemes and first-order convergence for the Euler scheme. 

\begin{table}[!htb]
  \centering
  \begin{tabular}{|l|c|c|c|c|}
    \hline
    Method & $K$ & $M$ & Rel.~Error & Time \\
    \hline
    Euler & $32$ & $2048000$ & $0.000934$ & $246.65$ sec \\
    \hline
    Ninomiya-Victoir & $4$ & $1024000$ & $0.002017$ & $52.33$ sec \\
    \hline
    NV with drift & $4$ & $1024000$ & $0.000862$ & $35.31$ sec \\
    \hline
  \end{tabular}
  \label{tab:gensabrmd-time}
  \caption{Computational time for the multi-dimensional generalized SABR model}
\end{table}

Further, the computational time for the generalized SABR model with a
four-dimensional stock market, reported in Table~\ref{tab:gensabrmd-time}, again
shows the usual picture. The classical Ninomiya-Victoir method gives a speed-up
between factors two and four (depending on the trust of the choice of $M$). In
this case, one might, however, note that the error from the classical
Ninomiya-Victoir method is more than twice higher the the errors from the
competing methods. And again, the simpler structure of the ODEs in the case of
a Ninomiya-Victoir method with drift leads to a convincing speed-up as
compared to both other methods.

\section*{Appendix}

In Sections \ref{example_heston}, \ref{sec:example:-sabr-model} and \ref{sv:gensabr}, the following ODE appears:
\begin{equation}
\label{powerODE}
\begin{split}
y'(t)= & h(t) (y(t))^\beta, \\
y(0)= & x,
\end{split}
\end{equation}
where  $0<\beta<1$, $x\geq0$ and $h:[0,\infty)\rightarrow\mathbb R$ is continuous and either positive valued or negative valued.
One can easily check that 
\begin{equation}
\label{solution}
\Phi_x(t)=\left( (1-\beta)  \int_0^t h(s)\mathrm{d}s +x^{1-\beta} \right)_+^{1/(1-\beta)}, \quad t\geq0, 
\end{equation}
with $a_+:=\max\{a,0\}$, is a solution to \eqref{powerODE}. We briefly  provide some details about uniqueness of the solution. When $h$ takes only negative values, then the right hand side of the ODE \eqref{powerODE} is decreasing in the state variable and uniqueness follows for any $x\geq0$ (see e.g. Example 2.4 on p.286 of \cite{karatzasshreve}). When $h$ takes only positive values and $x>0$, uniqueness follows by the Picard-Lindel\"of theorem. When $h$ takes only positive values and $x=0$, the solution \eqref{solution} is not unique; for instance, $y(t)\equiv 0$ forms another solution. For this particular case, we have chosen, throughout the paper, to work with the solution $\Phi_0(t)$, since the flow $\Phi_x(t)$ is (right)-continuous at $x=0$ for all $t\geq0$.

\end{document}